\documentclass[conference]{IEEEtran}
\IEEEoverridecommandlockouts
\usepackage{cite}
\usepackage{amsmath,amssymb,amsfonts}
\usepackage{algorithmic}
\usepackage{graphicx}
\usepackage{textcomp}
\usepackage{xcolor}
\usepackage[bottom]{footmisc}
\usepackage{url}

\def\BibTeX{{\rm B\kern-.05em{\sc i\kern-.025em b}\kern-.08em
    T\kern-.1667em\lower.7ex\hbox{E}\kern-.125emX}}

\begin{document}

\title{Graph-based Simulation Framework for Power Resilience Estimation and Enhancement\thanks{This material is based upon work supported by the Department of Energy, Solar Energy Technologies Office (SETO) Renewables Advancing Community Energy Resilience (RACER) program under Award Number DE-EE0010413. Any opinions, findings, conclusions, or recommendations expressed in this material are those of the authors and do not necessarily reflect the views of the Department of Energy.}}

\author{
\IEEEauthorblockN{Xuesong Wang, Shuo Yuan, Sharaf K. Magableh, Oraib Dawaghreh, Caisheng Wang*\thanks{*Corresponding author.}, and Le Yi Wang}
\IEEEauthorblockA{\textit{Department of Electrical and Computer Engineering} \\
\textit{Wayne State University}\\
Detroit, MI, USA \\
\{xswang, shuoyuan, sharaf.magableh, oraib.dawaghreh, cwang, lywang\}@wayne.edu}
}

\maketitle

\begin{abstract}
The increasing frequency of extreme weather events poses significant risks to power distribution systems, leading to widespread outages and severe economic and social consequences. This paper presents a novel simulation framework for assessing and enhancing the resilience of power distribution networks under such conditions. Resilience is estimated through Monte Carlo simulations, which simulate extreme weather scenarios and evaluate the impact on infrastructure fragility. Due to the proprietary nature of power network topology, a distribution network is synthesized using publicly available data. To generate the weather scenarios, an extreme weather generation method is developed. To enhance resilience, renewable resources such as solar panels and energy storage systems (batteries in this study) are incorporated. A customized Genetic Algorithm is proposed to determine the optimal locations and capacities for solar panels and battery installations, maximizing resilience while balancing cost constraints. Experiment results demonstrate that on a large-scale synthetic distribution network with more than 300,000 nodes and 300,000 edges, the proposed framework can efficiently evaluate the resilience, and enhance the resilience through the installations of distributed energy resources (DERs), providing utilities with valuable insights for community-level power system resilience estimation and enhancement.
\end{abstract}


\section{Introduction}

The resilience of power distribution systems is a critical area of study, particularly as extreme weather events become increasingly frequent and severe due to climate change. Notably, 80\% of major U.S. power outages reported from 2000 to 2023 were due to weather-related events \cite{R1}. Additionally, the number of weather-related power outages is increasing, with twice as many occurring during 2014-2023 compared to 2000-2009 \cite{R1}. Such events can result in prolonged power outages, causing significant economic losses and compromising the safety and well-being of communities. As a result, enhancing the resilience of power distribution networks has become a pressing concern for utilities, policymakers, and researchers. Effective resilience measures can help communities maintain reliable electricity amid extreme disruptions.

To study power system characteristics under disruptive events, two concepts, reliability and resilience, have been proposed \cite{R2}. Reliability typically addresses high-probability, low-impact events, whereas resilience focuses on low-probability, high-impact events \cite{R2}. 
%Resilience relates to the system's capability to maintain power supply during highly disruptive events, such as hurricanes, earthquakes, flooding, and cyber-physical attacks \cite{R2,R3}. 
There are two lines of study to assess the resilience of power systems, i.e., statistics-based methods and simulation-based methods. 

Statistics-based methods focus on a retrospective review of the disaster records, creating statistics metrics to describe the features of the target system. The Environment for Analysis of Geo-Located Energy Information (EAGLE-I) dataset provided comprehensive power outage records across counties in the United States from 2014 to 2022 \cite{R4}. Based on EAGLE-I data, the studies \cite{R5} and \cite{R6} proposed several metrics to measure the performance aspects of the power grid system during extreme outage events, such as event duration, impact duration, and impact level. A probability distribution function was obtained for each of these metrics. While \cite{R5} and \cite{R6} did not consider weather data, the study \cite{R7} combined the publicly available National Weather Service dataset with the EAGLE-I dataset to quantify power system resilience using three metrics: Time Over Threshold (TOT), Area Under the Curve (AUC), and Time After the End of the event (TAE). 

Simulation-based methods build models to emulate the physical systems, and test them under synthetic attacks, such as extreme weather and cyber attacks. It has the advantage of answering "what-if" questions. A quantitative framework for assessing microgrid resilience against windstorm is proposed in \cite{R8}, however, it is tested on a modified IEEE 33-bus distribution system which is far from real-world cases. A modeling framework is proposed to assess the resilience of power infrastructures under extreme weather events in \cite{R9}, however, similar to \cite{R8}, it is tested on a 6-bus transmission system. While resilience assessment methods share general procedures, most prior studies validated their frameworks on small \cite{R8,R9} or transmission systems \cite{R9}, lacking large-scale validation. Moreover, some of the existing research did not consider the joint study of power system resilience and enhancement \cite{R10,R11}.

Several methods to enhance power system resilience have been tested in prior research. For instance, line hardening strategies were used in \cite{R8,R9}. Additionally, mobile emergency generators are considered in \cite{R12} to be positioned before the disaster to hasten the likely post-disturbance operations. In \cite{R13}, a mobile energy resource allocation strategy is developed to restore critical loads by determining the most reliable restoration path against potential earthquakes. Besides the dynamic scheduling of mobile resources, static renewable resources, such as permanent solar or wind installations, are also considered in various literature. In \cite{R14}, a cooperative game theory-based approach is proposed to solve the sizing and siting problem of distributed energy resources. In \cite{R15}, an algorithmic formulation for optimal photovoltaic (PV) hosting and placement of distributed energy resources for network resiliency enhancement is developed using a Genetic Algorithm 
(GA), but tested on a 34-bus system. Similar to the works for resilience estimation, most of the work in resilience enhancement demonstrated their work in small-scale systems \cite{R12,R13,R14,R15}.

This study proposes a novel framework for resilience estimation and enhancement of large-scale distribution networks. Given a power system topology, an extreme weather model, and fragility models, the framework evaluates resilience by the trapezoid method \cite{R21} for each substation service area through Monte Carlo simulation. In the resilience enhancement phase, using the same topology together with damage and recovery records from the estimation phase, the framework determines the optimal locations and capacities of batteries and solar panels by using a customized GA. Several key function modules are designed to support the proposed framework. First, due to the proprietary nature of power network topologies, a synthetic distribution network is generated by using publicly available data, inspired by \cite{R16}. The network is modeled as a graph, with over 300,000 nodes and 300,000 edges. Different from the prior work that assumes an equal number of customers per building, this study estimates customer number distributions based on the building footprints, yielding a more realistic demand profile. Second, a thunderstorm wind scenario generator is developed by using public weather data. Third, a customized GA is designed to optimize a DER sizing and siting problem under cost constraints. To reduce the search space, a weighted location sampling strategy is introduced. To improve convergence stability, the tournament selection stage considers the best individuals not only from the current generation but also from all previous generations. Additionally, to promote spatial diversity in selected locations, a proximity rejection mechanism is implemented.

The main contributions of this paper are as follows:

\begin{itemize}
    \item A graph-based framework is proposed for the power system resilience estimation and enhancement that can handle large-scale distribution networks with over 300,000 nodes and 300,000 edges.
    \item A synthetic power grid topology generation method is proposed using publicly available data, including the estimated number of customers in each building.
    \item A weather scenario generator is developed that uses real-world data to simulate extreme wind events.
    \item A customized genetic algorithm is proposed to optimize the sizing and siting of batteries and solar panels, with weighted location sampling, global tournament selection, and proximity rejection mechanism.
\end{itemize}

% The remainder of this paper is organized as follows: Section II presents the methodology of the proposed framework. Section III covers the experimental setup and results, and Section IV summarizes key findings with the conclusion and discusses the limitations and future work.

\section{Methodology}

The proposed framework for resilience estimation and enhancement of power distribution networks consists of several key components: synthetic network generation, extreme weather scenario generation, fragility models, recovery strategy, Monte Carlo simulation for resilience estimation, and customized GA for DER siting and sizing. This section provides a detailed description of each component of the methodology.

\subsection{Synthetic Network Generation}

Inspired by \cite{R16}, we implement a synthetic distribution network generation method using publicly available data. The network is modeled as a graph, where nodes represent substations, poles, and buildings that contain customers, and edges represent distribution line segments. Three assumptions are made: 1) each customer is connected to one substation; 2) there is no connection between substations; 3) the power lines are along roads. The steps to build the network are as follows:

\begin{enumerate}
    \item Substation names are extracted from the reports published by the utility company, then cross referenced by using Open Street Map \cite{R17} to find their locations.
    \item Road data are extracted from the road maps\footnote{https://catalog.data.gov/organization/census-gov}. Line segments are created between consecutive points along the roads. Intermediate poles are created to make sure the distance between poles is not longer than 40 meters.
    \item Building information is extracted from the building footprint dataset\footnote{https://geo.btaa.org/}. For each building, the number of customers is estimated based on the residential or non-residential square footage.
    \item Each building and each substation are connected to their nearest poles. Each building is then assigned to a single substation via the shortest path, considering the power line length.
    \item To reduce the computation overhead and make the network more realistic, the buildings that are connected to the same pole on the same side of the line are merged into one node.
\end{enumerate}

\begin{figure}
    \centering
    \includegraphics[width=1\linewidth]{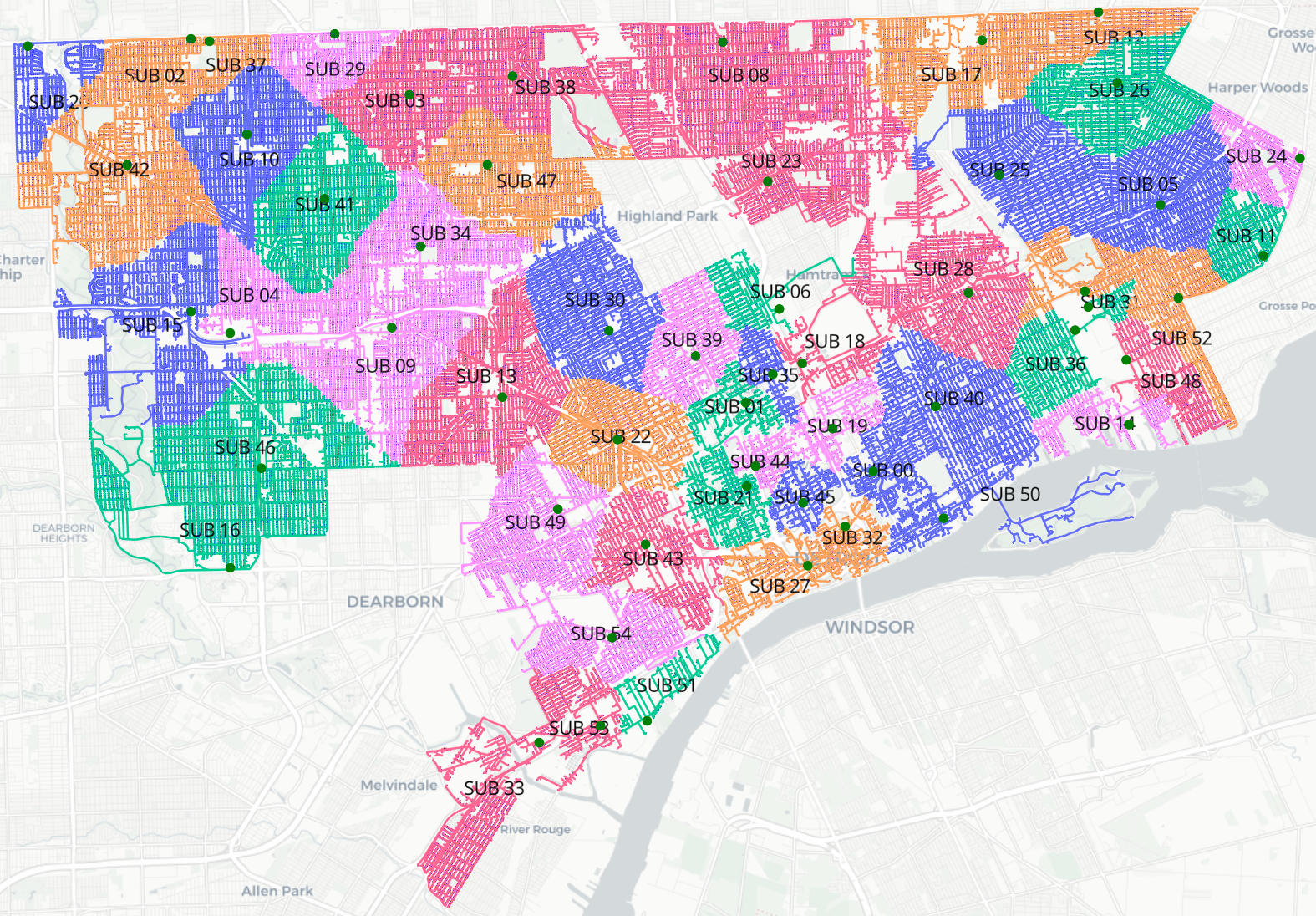}
    \caption{Synthetic distribution network of Detroit, MI.}
    \label{topology}
\end{figure}

This study uses Detroit, MI, as a case study area, with the generated topology shown in Fig.~\ref{topology}. Note that the proposed method can be extended to other areas as long as the public road and building footprint information is available.

\subsection{Extreme Weather Scenario Generation}

\begin{figure}[b]
    \centering
    \includegraphics[width=1\linewidth]{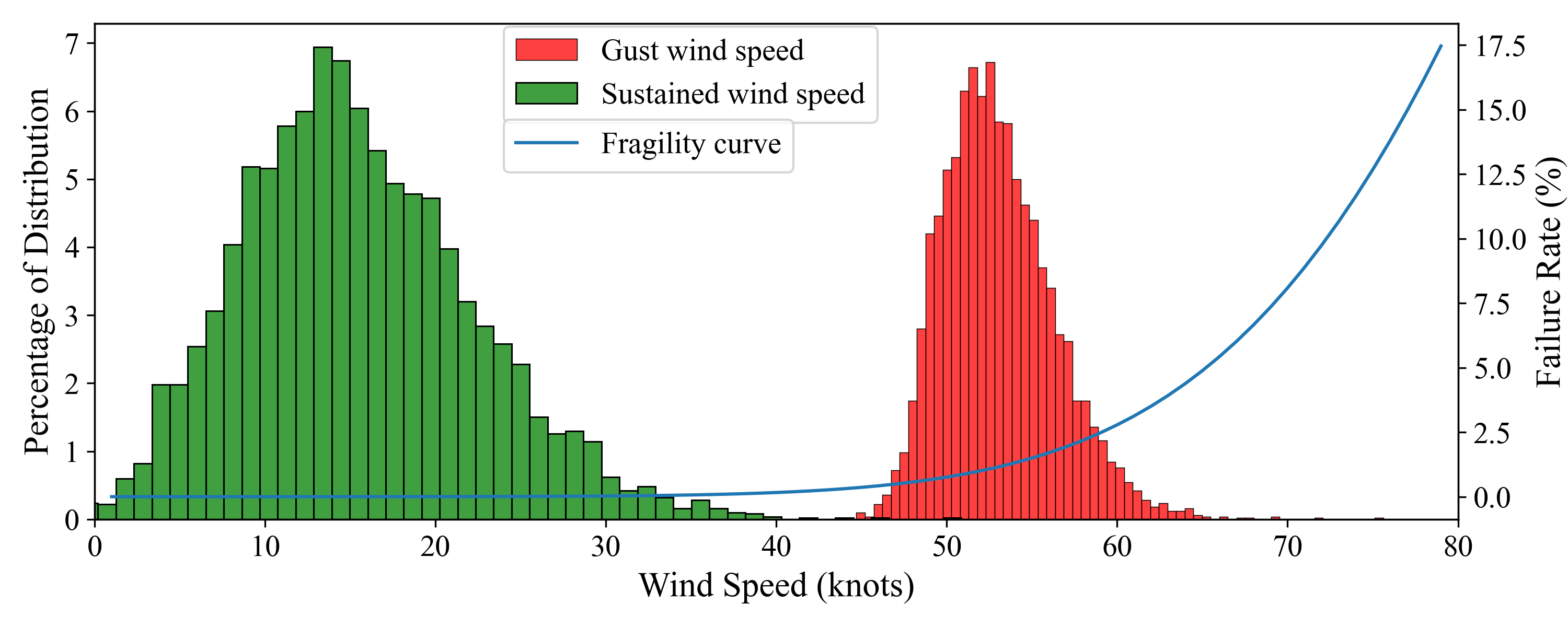}
    \caption{Wind speed distribution of gusts and sustained winds, along with the wind speed-based fragility curve.}
    \label{wind_frag}
\end{figure}

To simulate the impact of extreme weather events, we developed a weather scenario generator that produces spatial-temporal wind speed fields in thunderstorm events. The generator uses data from the NOAA Storm Events Database\footnote{https://www.ncdc.noaa.gov/stormevents/} and the HRRR model outputs \cite{R18,R19,R20} to capture real-world wind gust and sustained wind speed distributions for the studied area, respectively. Note that the proposed framework is highly scalable. Any extreme weather scenarios described in spatial-temporal field can be applied in this framework. The steps involved in generating the weather scenarios are:

\begin{enumerate}
    \item Wind gust and sustained wind speed distributions are fitted using log-normal distributions based on historical wind speed data of the studied area, as shown in Fig.~\ref{wind_frag}.
    \item The duration of storms is determined uniformly between 4 and 12 hours, with a proportion of hours experiencing wind gusts. The wind gust locations and timestamps are randomly selected.
    \item A sample point is uniformly drawn from each substation service area. The wind speed for each sample point is randomly drawn from the corresponding distribution, constituting a sparse wind speed field.
    \item The space of the studied area is segmented into patches, where each power system component belongs to one patch. The wind speed within the patch is assumed to be the same. The wind speeds of all the patches are calculated using 2-dimensional interpolation of the sparse field.
\end{enumerate}

\begin{figure}[b]
    \centering
    \includegraphics[width=0.7\linewidth]{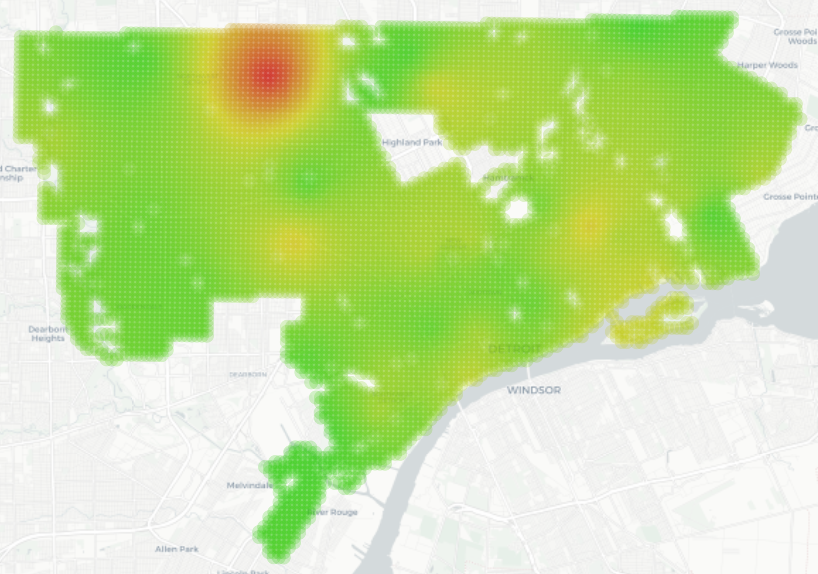}
    \caption{An example of the generated dense wind speed field, with wind gust locations shown in red and sustained winds in green.}
    \label{wind_speed}
\end{figure}

An example of the generated dense wind speed field is illustrated in Fig.~\ref{wind_speed}.

\subsection{Fragility Models}

\begin{figure}
    \centering
    \includegraphics[width=0.7\linewidth]{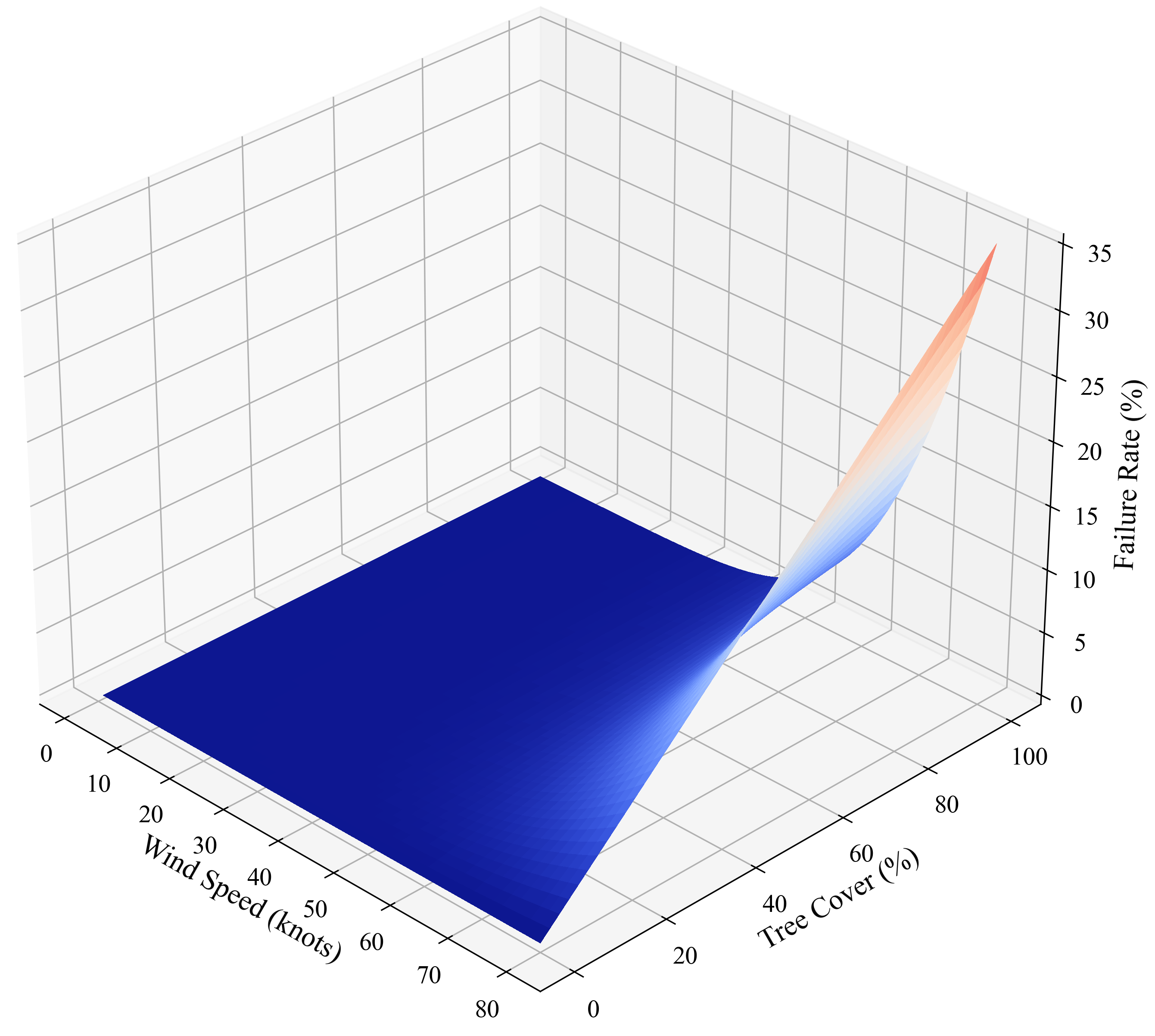}
    \caption{Fragility curve illustrating the joint influence of wind speed and tree coverage.}
    \label{wind_tree_frag}
\end{figure}

In this work, only fragility models of distribution lines are considered, with two failure modes analyzed:

\begin{itemize}
    \item Wind Speed-Based Fragility: This fragility curve represents the probability of line failure based on the local wind speed, as shown in Fig.~\ref{wind_frag}.
    \item Tree Coverage Influence: Tree coverage data obtained from USGS\footnote{https://data.fs.usda.gov/geodata/rastergateway/treecanopycover} is incorporated to determine the likelihood of line failure due to falling trees during wind events, as shown in Fig.~\ref{wind_tree_frag}.
\end{itemize}

\subsection{Recovery Strategy}

We implemented a criticality-based repair strategy, where crews prioritize repairing the line whose failure affects the most customers, calculated based on a criticality score. Multiple repair crews are considered, and repair times are uniformly randomized to reflect variability in field conditions. In this study, repair crews begin to work only after extreme weather conditions have subsided.

\subsection{Monte Carlo Simulation for Resilience Estimation}

To estimate resilience, we use the Monte Carlo method to simulate the impact of extreme weather events on the distribution network over multiple episodes. For each episode, a random extreme weather scenario is sampled and applied to the network hourly, and the power lines are broken based on the fragility models. After the extreme weather condition has subsided, the repair crews are dispatched to repair the broken power lines until all the power lines are repaired. For each episode, the resilience is measured by the trapezoid method \cite{R21}. Denote the resilience metric of area $i$ under episode $j$ by $R_{i,j}$. The final power resilience metric $R_i$ for substation service area $i$ is a weighted average of episode resilience values as shown in (\ref{resilience_eq}). A higher weight is given to the episode when a wind gust occurs in the substation service area, $\lambda$ is the weighting term. $N_{g,i}$ is the number of episodes with a wind gust in the substation service area $i$, and $N$ is the total number of episodes.

\begin{equation}\label{resilience_eq}
    R_i=\frac{\lambda}{N_{g,i}}\sum_{j=1}^{N_{g,i}}{R_{i,j}} + \frac{1-\lambda}{N-N_{g,i}}\sum_{k=1}^{N-N_{g,i}}{R_{i,k}}
\end{equation}

\subsection{Customized Genetic Algorithm for DER Siting and Sizing}

To search for the optimal locations and sites of the solar panels and batteries, a customized GA is proposed. The GA evolves a population of candidate solutions over generations, where each individual is encoded as a chromosome representing the independent variables, i.e., the locations and sizes of the solar panels and batteries. The algorithm performs selection, crossover, and mutation to generate new solutions, favoring those with higher fitness scores. The constraints of the problem are the total number of locations and the capacity limits of the generators at each location. The optimization goal is to maximize the lowest resilience score over all the substation service areas.

Because the extreme weather considered in this work is the wind hazards during thunderstorms, solar panels operate only after the weather scenario ends and when they are in island mode, while the batteries work all the time as long as they are in island mode and still have energy. The customized GA proposes candidate solutions in each generation. Then, each solution is installed in the network, and all the episodes from the estimation phase are replayed, considering the effects of solar panels and batteries. The lowest resilience score over all the substation service areas is reported as the fitness score of that solution.

In order to adapt the GA to this context, several modifications have been made to the traditional GA:

\begin{itemize}
    \item Global tournament selection: In the selection stage, if the best individuals are selected from the current generation only, the algorithm is not convergent. To improve convergence, the best individuals are selected from the current generation and all previous generations.
    \item Proximity rejection: When generating new individuals, it is imperative to select spatially diverse locations. Locations are selected one at a time. When a new location is selected, the shortest distance to the nearest location among all the selected locations is denoted as K (hops). Then, the newly selected location is rejected with a probability of 1/K . It is also applied in the crossover step.
    \item Weighted location sampling: The number of episodes in which the node loses power in the estimation phase is counted, then the numbers are scaled by the node degree. Finally, the numbers are normalized to produce a probability distribution.
\end{itemize}

\section{Experiment Results and Analysis} 

\subsection{Power Resilience Estimation}

\begin{figure}[b]
    \centering
    \includegraphics[width=0.7\linewidth]{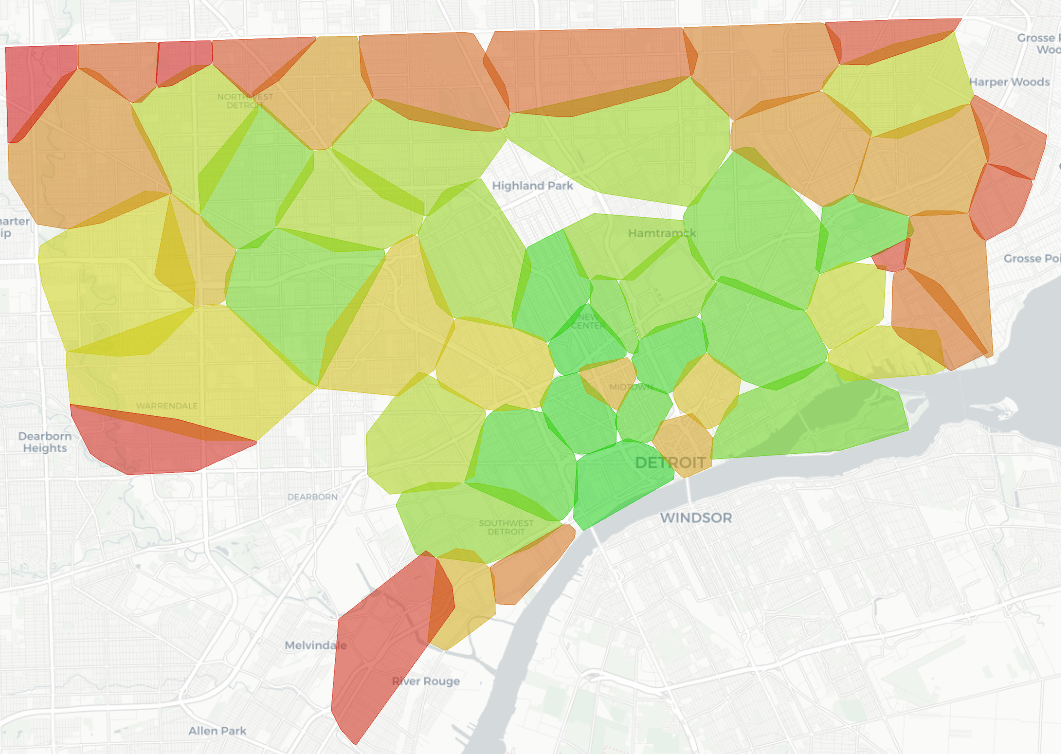}
    \caption{Resilience estimation results across 10,000 episodes, where red denotes lower resilience (higher vulnerability) and green denotes higher resilience.}
    \label{resilience_estimation}
\end{figure}

\begin{figure}
    \centering
    \includegraphics[width=0.9\linewidth]{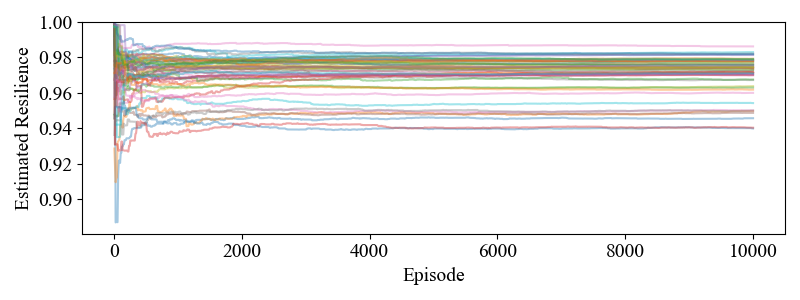}
    \caption{Convergence analysis of all the substation service areas.}
    \label{convergence}
\end{figure}

The Monte Carlo simulations were run for $N=10,000$ episodes. $\lambda$ is set to 0.8 in this work. The weighting term $\lambda$ controls the importance of the wind gust scenarios in the studied area, as wind gusts have a higher impact on the system compared to sustained winds. It can be adjusted by framework users to align with real-world considerations. The resilience estimation results are displayed in Fig.~\ref{resilience_estimation}. The convergence analysis displayed in Fig.~\ref{convergence} showed that the resilience metrics converged as the number of episodes increased, indicating that the simulations provided statistically reliable estimates of network resilience. This convergence is critical for ensuring that the resilience scores obtained from the simulations are representative of the network's performance under a wide range of extreme weather scenarios.

\subsection{Renewable Resource Integration and Resilience Enhancement}

\begin{figure}
    \centering
    \includegraphics[width=0.85\linewidth]{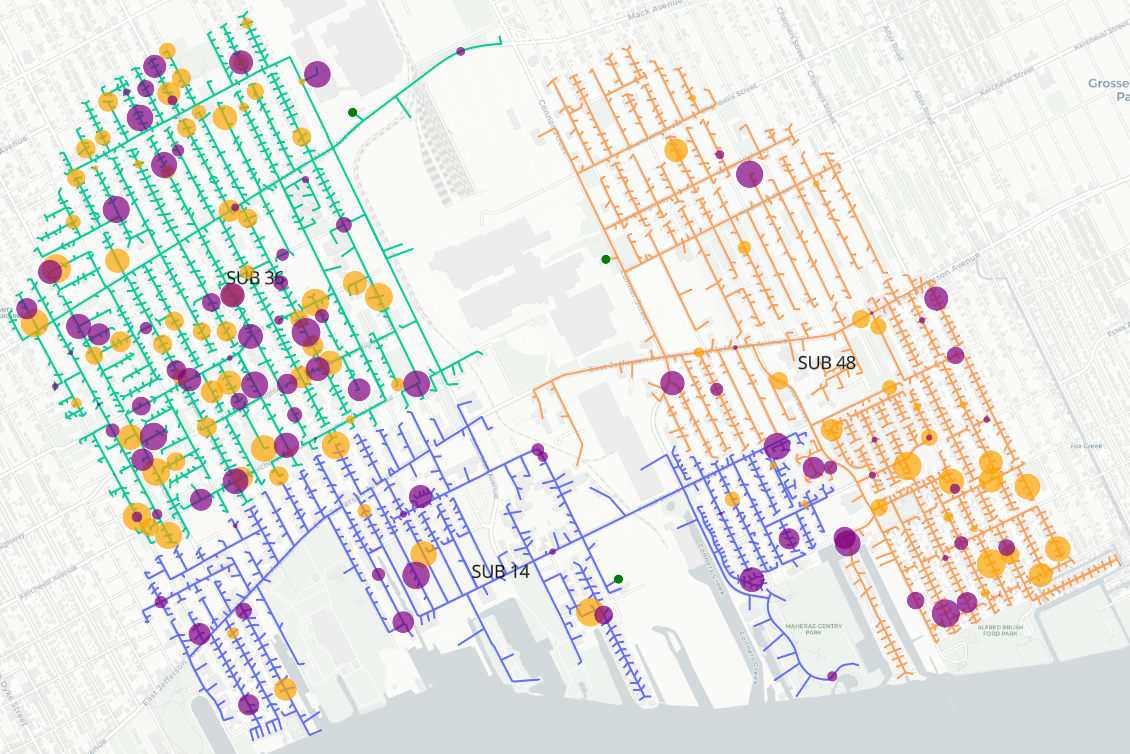}
    \caption{Optimal sizes and locations of solar panels and batteries in three substation service areas ($SUB14$, $SUB36$, and $SUB48$). Purple dots represent batteries, and yellow dots represent solar panels. The size of the dot is proportionate to the installed capacity.}
    \label{best_plan_searched}
\end{figure}

\begin{figure}
    \centering
    \includegraphics[width=1\linewidth]{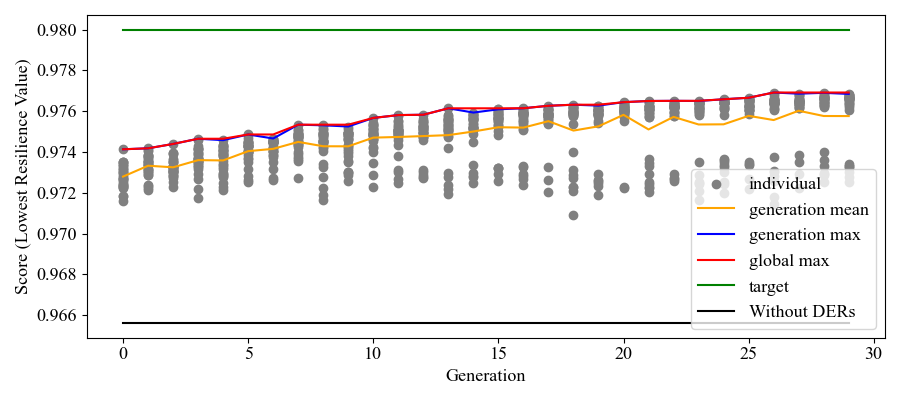}
    \caption{Score progression of the customized GA over generations.}
    \label{plan_history}
\end{figure}

Three substation service areas are selected to demonstrate the optimal sites and sizes of the solar panels and batteries, i.e., $SUB14$, $SUB36$, and $SUB48$. The results are shown in Fig.~\ref{best_plan_searched}. Fig.~\ref{plan_history} shows the score progression of the customized GA. As can be seen from the figure, the algorithm can consistently search better solutions in almost every generation, illustrating the efficacy of the proposed customized GA.

\section{Conclusion}

This work presented a framework for assessing and enhancing the resilience of power distribution networks under extreme weather conditions. The graph-based model, combined with Monte Carlo simulations, allowed for quantitative analysis of network resilience. DERs, specifically solar panels and batteries, were integrated to enhance resilience, with optimal placement and sizing determined by a customized GA. The effectiveness of the proposed framework was demonstrated using a case study of the Detroit area. The results showed that the framework can estimate power resilience for large-scale distribution network and consistently search for optimal siting and sizing solutions for renewable integration. Future work includes exploring multiple types of extreme weather events, evaluating diverse recovery strategies, incorporating additional failure modes, conducting joint analyses with other critical infrastructures, and validating the simulation results using real-world data.

\bibliographystyle{ieeetr}
\bibliography{main}

\end{document}